\documentclass{article}

\usepackage{arxiv}
 \usepackage{latexsym,amsfonts}
\usepackage{graphicx}
\usepackage{algorithm}
\usepackage{algorithmic}
\usepackage{multirow}
\usepackage{amssymb}
\usepackage{mathrsfs}
\usepackage{amsmath}
\usepackage{amssymb}
\usepackage{mathrsfs}
\usepackage{CJK}
\usepackage{multicol} 
\usepackage{hyperref}

\newcommand{\beit}{\begin{itemize}}
\newcommand{\eit}{\end{itemize}}

\newsavebox{\savepar}

\def\setsize{\csname @setfontsize\endcsname \setsize}

\begin{document}
\title{A fine-grained policy model for Provenance-based Access Control and Policy Algebras}
\author{Xinyu Fan\thanks{Corresponding author}, Faen Zhang  Jianfei Song Jingming Guo Fujie Gao\\
fanxinyu@ainnovation.com \{zhangfaenainnovation, songjianfeiainnovation\}, \\
\{guojingmingainnovation, gaofujieainnovation\}@gmail.com\\
AInnovation Technology Ltd.} 
\maketitle              
\begin{abstract}
	A fine-grained provenance-based access control policy model is proposed in this paper, in order to improve the express performance of existing model. This method employs provenance as conditions to determine whether a piece of data can be accessed because historical operations performed on data could reveal clues about its sensitivity and vulnerability. Particularly, our proposed work provides a four-valued decision set which allows showing status to match a restriction particularly. This framework consists of target policy, access control policy, and policy algebras. With the complete definition and algebra system construction, a practical fine-grained access control policy model is developed.

\end{abstract}
\section{Introduction}

Data provenance logs historical operations performed on documents to protect their security and privacy. Provenance Access Control is considered to be an important research topic of big data security. The sensitivity of files and their provenance may vary, and users may request and be granted access to files and provenance separately. In some cases, provenance itself may contain sensitive information that may require more protection than the attached document. For instance, while a programming project can be published to the public, its authors and operations should be kept secret to prevent technology leaks. Therefore, access control is required  for the provenance data itself. It allows qualified users to access provenance data and protects it from unauthorized access.  Provenance can be expressed as a directed acyclic graph (DAG),  illustrating how the data artifacts are processed. In such a provenance DAG under the Open Provenance Model (OPM)\cite{MoreauCFFGGKMMMPSSB11}, nodes represent three main entities, including \emph{Artifact}, \emph{Agent} and \emph{Process}, and edges represent connections to the main entities.

Several provenance-based access control policy models \cite{NguyenPS13}\cite{ParkNS12} and history-based access control \cite{NiBL09} have been proposed. Park \emph{et al.}\cite{ParkNS12} proposed a family of provenance-based access control models which utilised a notion of dependency as the key foundation for access control policy specification. They proposed a basic provenance-based access control model $PBAC_B$, which facilitates additional capabilities beyond those available in traditional access control models by introducing provenance as access control conditions. Based on the basic model $PBAC_B$, a family of PBAC models was defined by extending three criteria, which are (1) the kind of provenance data in the system, (2) whether policies are based on acting user dependencies and object dependencies, and (3) whether the policies are readily available or need to be retrieved. The three models $PBAC_U$, $PBAC_A$ and $PBAC_{PR}$ extend one of these three criteria respectively. However, combined results for individual policies are only basic conjunctive or disjunctive connectives between rules. 

However, previous provenance-based access control policies have not solved some problems. The research of partial matching of provenance map is not deep enough so that the existing work does not implement the appropriate fine grained strategy model. The research of partial matching of provenance map is not deep enough.  Partial matching a provenance graph is a scenario where  not all elements of a provenance subgraph can be found in the given provenance graph. For example, let a policy condition be: \emph{a doctor diagnosed a patient at 22/4/2018}. The partial matching scenario could be that a doctor diagnosing a patient, but the timestamp is not 22/4/2018 or missing. It is important to propose a mechanism to introduce partial matching provenance  graph, and values showing the partial matching state should be given in a fine-grained policy model. From this point, policies algebras to merge those partial matching values are also missing from the existing work. 

To tackle the problems mentioned above, we propose a framework based on OPM+, which  is composed of three types of atomic targets policies proposed by us. Path atomic target policy only takes attributes from provenance. The determination is based on whether certain operations that are recorded in provenance have been performed on the data. Associated atomic target policy takes attributes from both provenance and requests (subject, object, \emph{etc.}), which implies that the determination is based on whether the requesters have performed operations on the data. The latter atomic targets perform differently from traditional access control. For instance, an associated atomic target could express such a scenario: if the requester was editing a file before and have not submitted it, the access is permitted. Notably, for this scenario, the ``requester" is the subject of a query, and the editing and submitting are processes from a provenance. At last, the attributes from requests can also be defined as conditions, which is named as request atomic target. A fine-grained provenance-based access control framework consists of policy models, policy evaluation, and policy algebras. The contributions can be summarized as follows:
\begin{itemize}
	\item We propose three types of atomic targets policies: path atomic target policy, associated atomic target policy and request atomic target. 
	\item We define a ``four-valued" decision set for each atomic target (conditions). The attribute triple is replaced as a provenance partition, which can express a series of historical operations. 
	\item We rearrange logic operators and create several new operators to combine atomic policies.
\end{itemize}

\section{System Assumption}

\begin{figure}[thb]
	\vspace{-0cm}
	\centering
	\includegraphics[scale=0.3]{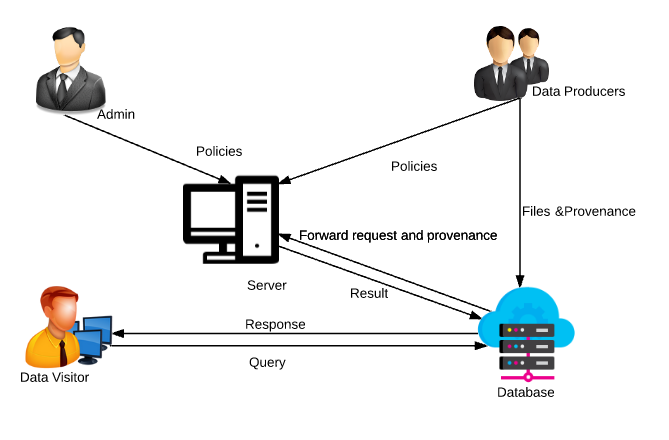}
	\vspace{-0.5cm}
	\caption{System Assumption} \label{fig11}
\end{figure}

To establish the framework of provenance-based access control model and policy algebras, we propose a system assumption shown in Fig\ref{fig11}, where the system consists of four parts including \emph{Administrator}, \emph{Server}, \emph{Data Producers} \emph{Data Visitors} and \emph{Data Storage}, which we will illustrate in the follwing.

\begin{itemize}
	\item \emph{Administrator} generates provenance-based access control policies and sends them to the \emph{Server}.
	\item \emph{Server} collects policies from administrators and (optional) data producers. The server evaluates results based on the policies and delivers the results to databases. 
	\item \emph{Data Producers} generate files and attached provenance which are sent to \emph{Data Storage}. They might also produce self-defined access control policies and send them to the \emph{Server}.
	\item \emph{Data Visitors} send queries to the \emph{Database} in order to access files stored in the database. 
	\item \emph{Data Storage} stores files and provenance, which receives queries from users and responds to queries according to the results generated by the \emph{Server}.  
\end{itemize}

In a provenance-aware system, access control policies can be generated by system administrators and data owners. When a data visitor sends a request to a server to access a piece of data, the server needs to retrieve the provenance of targeted data as the policies are tailored to make access decisions based on data provenance. After receiving results from the server, the database implements the results by approving or rejecting users' requests.  It can be noticed that most access control mechanisms take evaluations based attributes from queries or current system conditions. While for provenance-based access control policies, conditions and restrictions can be extended to provenance which logs historical operations of data. 

\section{Target Policies}

The provenance-based access control framework consists of access control policies and policy algebras, where each policy consists of target section and access control section. Each provenance-based access control policy consists of a target section and an access control section. The \emph{target} section confines applicable requests of each policy. The policy can affect the request, only if the objective provenance graphs in a request meet the conditions of \emph{Target}. In addition, the access control section evaluates the accessibility of the request. We give definitions of syntax for targets. 

\begin{figure}[thb]
	\begin{minipage}[t]{0.5\linewidth}
		\vspace{-0cm}
		\centering
		\includegraphics[scale=0.25]{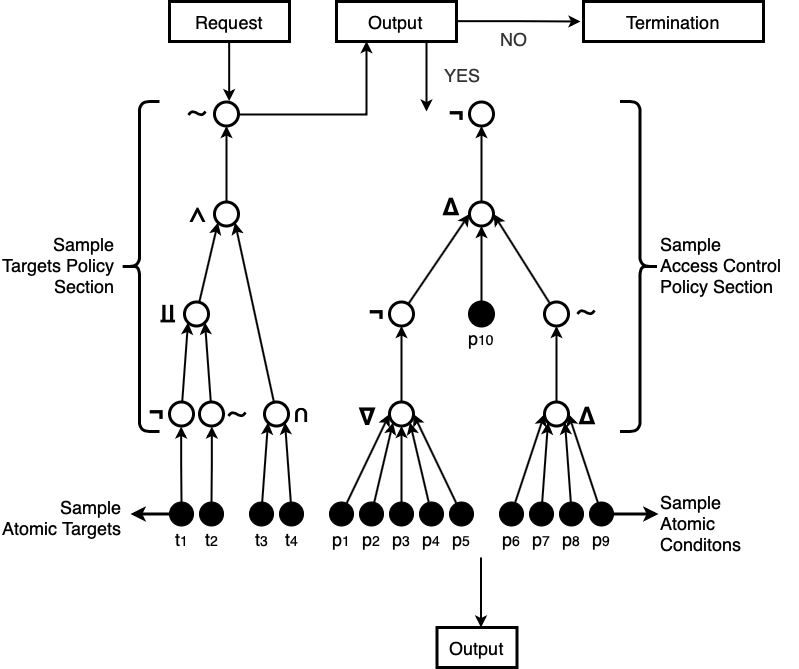}
		\vspace{-0cm}
		\caption{The framework of Proposed Provenance-based Access Control} \label{fig1}
	\end{minipage}
	\begin{minipage}[t]{0.5\linewidth}
		\vspace{-0cm}
		\centering
		\includegraphics[scale=0.3]{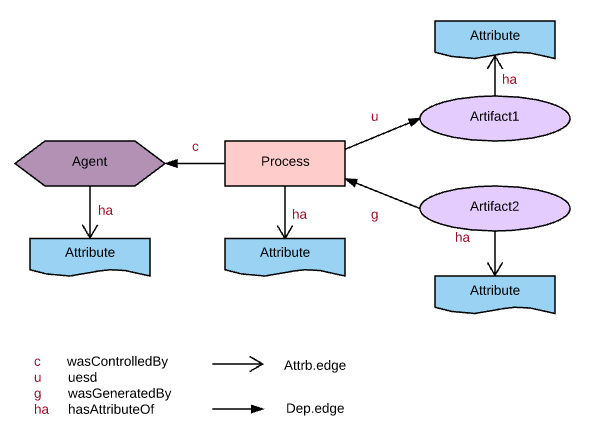}
		\vspace{1cm}
		
		\caption{OPM$^+$ Schema} \label{fig1}
	\end{minipage}
	
\end{figure}

{\bf Definition 1} (Open Provenance Model$^+$(OPM$^+$)) is an extension of OPM, which records how is a piece of data derived. The model is defined by a triple $<$T, L, G$>$:

\begin{itemize}
	
	\item T is the vertex types: agent (Ag), artifact (A), process (P) and attribute (Att). Each vertex in a provenance graph is one of these types. In Figure 2, an artifact is represented by the shape of oval, which is an object or as a piece of data, such as ``$homework_1$", ``\emph{comments}", and \emph{etc.}; a process is an action which is the operation executed on a piece of data, such as ``\emph{submit}" and ``\emph{review}"; an agent is a subject that sponsors an action including ``\emph{$user_1$}" and ``\emph{professor}".   
	
	\item L is the relationship labels: used (u), wasGeneratedBy (wgb), wasControlledBy (wcb), wasTriggeredBy (wtb), wasDerivedFrom(wdf) and hasAttributes (ha).  Each edge in a provenance graph will be labeled as one of these labels. The Labels describe the relationships between the vertices.
	
	\item G is a labelled DAG, where G = $<$V, E$>$, E defines the allowable relationships between the elements, E = \{ (P, A, used), (A, P, wgb), (P, Ag, wcb), (A, A, wdf), (P, P, wtb), (Ag, Att, ha), (P, Att, ha), (A, Att, ha) \}
	
\end{itemize}

Given the OPM$^+$ $<$T, L, G$>$, an OPM$^+$ instance is defined by a provenance graph $G_i$ = $<V_i, E_i>$, where $V_i$ is a set of entities and $E_i$ $\in V_i \times V_i \times L$. Let $\tau$: $V_i \rightarrow T$ be a function that maps an entity to its type, we say $G_i$ is valid if for each entity v $\in V_i$, $\tau(v) \in$ T, and for each edge (v, v$^\prime$, l) $\in E_i$, ($\tau$ (v), $\tau$ (v$^\prime$), l)$\in$ E. We extend the definition of OPM in paper \cite{ChenENN15}. \\

{\bf Definition 2} (Attribute Node) The extended provenance model attaches context information as attribute nodes in DAG, where each attribute node consists of attribute items and attribute values. For instance, let $Attribute_i$ for a process be \{timestamp: 1/5/2017; system condition: Linux; location: Sydney\}. In this model, provenance can be classified into base provenance data and (optional) attribute provenance data that is associated with main entities in the graph (ag, p, or a). Attribute provenance data is classified into three categories: Agent-related Attributes, Process-related Attributes, and Artifact-related Attributes.  

\begin{itemize}
	
	\item{Agent-related Attributes:} Agents trigger and execute operations, and their attributes include IDs, activated roles \emph{ect.} Identities are usually unique labels to identify users. Activated roles are the roles users employ by taking actions and can play a key role in distinguishing the sensitivity of content and in making access decisions. For instance, when Alice adds a piece of data into a document in the role of teaching assistant, she grades assignments of students. The comments and scores can only be accessed by Alice and the owner of assignments. While when Alice edited data with a role of student, the content of assignments can be read by other students enrolled the same subject. 
	
	\item{Process-related Attributes:} Processes are operations performed on data and result in the change of data, their related attributes include temporal aspects when operations are performed, such as locations, timestamps, system conditions \emph{etc.} They can influence access decisions. For example, operations in provenance can be accessed if they were executed before 2016.  
	
	\item{Artifact-related Attributes:} Artifacts are objects including input messages, output messages, and source data. The related attributes can include object size, permitted usages defined by the data producers \emph{etc.} Some meta-data that might normally be recorded as attributes may not be held in this way in provenance data, as they are recorded within the structure of a provenance graph. For example the generating agent and time of data. 
	
	\end {itemize}

	{\bf Definition 3} (Path Atomic Target). We define four types of atomic targets from provenance graphs: \\
	-$null_T$ is a target;\\
	-$(v_{type}, v_{name})$ is a target, where $v_{type}$ is a vertices type and $v_{name}$ is a vertices name;\\
	-$(v_{type}, v_{name}, a, f)$ is a target, where $v$ is a vertices, $a$ is an attribute value and $f$ is a binary predicate. \\
	-a target is a string of $(v_{type}, v_{name})$ or $(v_{type}, v_{name}, a, f)$ expressed over XPath.\\
	-($v_{type}, v_{name}$, $x$, $f$) is an alternative type of atomic target, where ($v_{type}, v_{name}$) is a vertices, $x$ is an attribute in a query, and $f$ is a binary predicate.
	
	First, the atomic target models are named as \emph{path atomic target}, as they extract elements from provenance. A signal vertex can be viewed as an extreme form of a provenance path.  An atomic target could be one or a string of quaternions, which illustrates a dependent path. $v_{type}$ is the type of a vertex, which is one element in the set of V$^*$ = \{Ag $\cup$ A $\cup$ P\}. 
	
	$v_{name}$ is an element in a finite set of abstracted names for vertices, where an abstracted name expresses a class of vertices names. For instance, \emph{User} is the abstracted name for all the users including \emph{$user_1$}, \emph{$user_2$} \emph{etc.} In terms of the motivation for using abstract value names in policies, provenance is captured after provenance-based access control policies are defined, and provenance evolves as the file grows. Therefore, the exact names of vertices can be infinite and unpredictable. Therefore, it is difficult for policy generators to predict specific names for all vertices, and we use abstract vertex names in provenance-aware policies. In provenance-aware systems, provenance graphs can be interpreted as Typed Provenance Model (TPM) \cite{SunP0S16} for policy checking purposes,which bridges the gap between policies and provenance graphs. In TPM, vertex names are interpreted as abstract names or vertex class names.
	
	\emph{a} is an attribute for vertices. The extended provenance model OPM$^+$ stores context information as attribute nodes in a provenance DAG. Under this model, provenance can be divided into basic provenance data and (optional) attribute provenance data associated with main entities in the graph  (ag, p, or a). Attribute provenance data is divided into three categories: Agent-related Attributes, Process-related Attributes, and Artifact-related Attributes. 
	
	In quaternion Qua = $(v_{type}, v_{name}, a, f)$, \emph{f} is a binary predictor. 
	Binary predictors have five possible values: $\leqslant$, $<$, $\geqslant$, $>$. Also, since equality (=) is adopted most frequently, we assume equality is the default value and can be omitted from the atomic targets. For example, let a quaternion be (\emph{Agent}, \emph{User}, \emph{Female}, =). In the example, \emph{Agent} is the vertices type; \emph{User} is the abstracted vertice's name in a graph; \emph{Female} is an attribute for the vertices; and = is the binary predicate. Namely, a user vertices of type \emph{Agent} with female attribute matches this atomic target. For another example, (\emph{Process, Submit/wasSubmittedBy}, \emph{1/1/2016}, $<$) is a quaternion, and a process named ``Submit" or ``wasSubmittedBy" executed before \emph{1/1/2016} matches it. 
	
	A dependency  path can be not only a single vertex in the source graph, but also a string of connected vertices, which represent a set of operations performed on a piece of data.  Whether a provenance is composed of given dependency paths can be defined as a condition in the access control policy, since provenance-based access control determines whether a piece of data is accessible based on whether certain operations are performed on the data. We propose a few simple samples to illustrate what provenance-based access control policies can do:
	
	(1) \emph{An exam paper can be downloaded if it has been reviewed at least three times but not graded;}
	
	(2) \emph{A paper can be accessed if it has not been submitted;}
	
	(3) \emph{A file can be accessed if it was generated originally before 2016 and was reviewed after 2017;}
	
	(4) \emph{If a piece of data was revised by Alice in 2016, and never been edited by Bob, then it can be accessed by students;}
	
	The examples above are sample provenance-based access control policies that make access decisions based on certain operations recorded in provenance. One or a series of operations can be defined as a provenance path in policies. Particularly, in certain scenarios, a predefined dependency path  may list only a few vertices in a string, such as the starting point and ending point of a provenance path. Thus, it solves problems where it is difficult to predict all vertices in a provenance path. For example, let a dependency path be ($v_i, \backslash$v+, $v_j, \backslash$v\{2\}, $v_k$). It presents a path starting at vertices $v_i$, followed by several other vertices, regardless of the number of them and of vertices $v_j$. The path ends at $v_k$, where there are two vertices between $v_j$ and $v_k$. 
	
	\emph {Directed $D_{PATH}$} is connected by all types with the \emph{CT}, where the processes in a path are presented in a timing sequence. Thus, by identifying the directions of labels that connect the vertices, it ensures that operations are executed according to time order in a directed $D_{PATH}$.

	\subsection{Policy Evaluation}
	With respect to the evaluation of the atomic target relative to the request, the atomic target matches the query if the provenance graph satisfies all the attributes in the atomic target. Otherwise, it does not match the atomic target. 
	
	We start with the \emph{request atomic target}, as it takes the form as a single value. When the attribute in a query satisfies the value of attributes, the result is $1_T$; otherwise, the result is $\times_T$.  
	
	Because the model of \emph{path atomic target} and \emph{joint atomic target} adopt the same form. Together we explained their evaluation principles. We hope there exist values of a result to represent the conditions between a full match or complete not match. Hence, we present a four-valued result set $Dec_T$ = \{$1_T, 0_T, \perp_T, \times_T$\} for the evaluation, where $1_T$ is absolute match and $\times_T$ represents completely not match. There is $1_T > 0_T > \perp_T > \times_T$.
	
	(1) The \emph{``null"} target matches all requests, which implies it returns $1_T$ for all objective provenance graphs.
	
	(2) If the provenance graph contains vertices with the vertex type and name, the target ($v_{type}$, $v_{name}$) matches, and the vertex type and name return the value of $1_T$. 
	However, when $v_{name}$ does not match, it returns $\perp_T$; When neither matches, it returns $\times_T$.
	
	(3) When the quaternion ($v_{type}$, $v_{name}$, $a$, $f$) or ($v_{type}$, $v_{name}$, $x$, $f$) is met, it returns $1_T$; if it only matches $v_{type}$ and none of the other three values match, it returns $0_T$. If the provenance graph only meets ($v_{type}$, $v_{name}$) but not the value, it returns $\perp_T$; $\times_T$ denotes that it does not contain the vertex $v$ in the graphs.
	
	(4) $D_{path}$ is quaternion of ($v_{type}$, $v_{name}$, $a$, $f$) or ($v_{type}$, $v_{name}$, $x$, $f$) defined by a string of nodes, when it meets all quaternion, it returns $1_T$; It returns $0_T$ if it meets all the $v_{type}$, $v_{name}$ in the string but not the other two values; If only all $v_{type}$ is satisfied, it returns $\perp_T$ ; It returns $\times_T$ when the provenance DAG can not meet all the $v_{type}$ in the string.
	
	We denote the evaluation of a given request \emph{q} as $[\![t]\!]_T$($q$) $\in$ $Dec_T$. We utilise the subscript $T$ to indicate that it is an evaluation of the target to distinguish the results of target section from that of the access control section. However, the subscript can be issued without ambiguity. Initially, a query for \emph{null} target returns $1_T$, and an evaluation for a query with empty provenance graph returns $\times_T$.  $[\![$ null $]\!]$($q$)=$1_T$;         
	$[\![$n$]\!]$($\varnothing$)=$\times_T$
	
	A more fine-grained provenance-based access control policy model is proposed by defining and evaluating atomic targets. In this model, different types of attributes extracted from a provenance graph are distinguished. Namely, the four values in the decision set represent the case where a provenance graph satisfies different categories of elements in the atomic target. 
	
	\subsection{Operators}
	A target section may consist of several atomic targets organized under a tree structure. In a tree structure, leaf nodes are atomic targets and non-leaf nodes are logic operators that combine the results of each atomic target to output a final decision.
	
	The operators are tailored for our proposed four-valued decisions set, including binary operators, unary operators, and u-ary operators. We introduce binary and unary operators \cite{CramptonM12}, which correspond to the weak and strong Kleene operators \cite{KleeneS50} of the four values target decision set \{$1_T$, $0_T$, $\perp_T$, $\times_T$\}, and define several new operators. We define operators $\bf{not}$ $t$, $\bf{opt}$ $t$, $t_1$ $\bf{and}$ $t_2$ and $t_1$ $\bf{or}$ $t_2$ of targets as follows:
	
	\begin{center}
		$[\![$ $\bf{not}$ $t$ $]\!]$($q$)= $\lnot$ $[\![$ $t$ $]\!]$($q$);  
		$[\![$ $t_1$ $\bf{and}$ $t_2$ $]\!]$($q$)=$[\![t_1]\!]$($q$) $\sqcap$ $[\![t_2]\!]$($q$)\\ 
		$[\![$ $\bf{opt}$ $t$ $]\!]$($q$)= $\sim$ $[\![$ $t$ $]\!]$($q$);  
		$[\![$ $t_1$ $\bf{or}$ $t_2$ $]\!]$($q$)=$[\![t_1]\!]$($q$) $\sqcup$ $[\![t_2]\!]$($q$)
	\end{center}
	
	Here, $\bf{not}$ t and $\bf{opt}$ t are unary operators. The total order of $Dec_T$ is that $1_T$ $>$ $0_T$ $>$ $\perp_T$ $>$ $\times_T$. In binary operator $\bf{and}$, the result is the lower value between the two sides. To the contrary, $\bf{or}$ outputs the higher values from the two inputs. However, if we change the $Dec_T$ order as $\times_T$ $>$ $\perp_T>0_T>1_T$, $\bf{and}$ (in $1_T$ $>$ $0_T$ $>$ $\perp_T$ $>$ $\times_T$)= $\bf{or}$ (in $\times_T$ $>$ $\perp_T>0_T>1_T$).\\
	\begin{minipage}{\textwidth}
		\begin{minipage}[t]{0.45\textwidth}
			\centering
			\makeatletter\def\@captype{table}\makeatother
			\begin{tabular}{c|c|c|c}
				X & $\lnot$X & $\sim$X & $\star$X\\
				\hline
				$1_T$ & $0_T$ & $1_T$ & $\perp_T$\\
				$0_T$ & $1_T$ & $\perp_T$ & $\times_T$ \\
				$\perp_T$ & $\perp_T$ & $0_T$ & $1_T$ \\
				$\times_T$ & $\times_T$ & $\times_T$ & $0_T$\\
			\end{tabular}
			\caption{Unary Operators}
		\end{minipage}
		\begin{minipage}[t]{0.45\textwidth}
			\centering
			\makeatletter\def\@captype{table}\makeatother
			\begin{tabular}{c|c c c c}
				$\sqcup$ & $1_T$ & $0_T$ & $\perp_T$ & $\times_T$\\
				\hline
				$1_T$ & $1_T$ & $1_T$ & $1_T$ & $1_T$\\
				$0_T$ & $1_T$ & $0_T$ & $0_T$ & $0_T$\\
				$\perp_T$ & $1_T$ & $0_T$ & $\perp_T$ & $\perp_T$\\
				$\times_T$ & $1_T$ & $0_T$ & $\perp_T$ & $\times_T$\\
			\end{tabular}
			\caption{Binary Operator $\sqcup$}
			
		\end{minipage}
		
	\end{minipage}

	Hence, when we change the priority order for the values in the $Dec_T$ set, we can enumerate all the operators as $\bf{and}$ for the orders $1_T>0_T>\times_T>\perp_T>$ ($\cup$), $1_T>\perp_T>0_T>\times_T$ ($\cap$), $1_T>\times_T>\perp_T>0$ ($\sqsubset$), $0_T>1_T>\perp_T>\times_T$($\wedge$), $0_T>\times_T>\perp_T>1_T$($\vee$), $\times_T>1_T>0_T>\perp_T$($\supset$), $\times_T>\perp_T>0_T>1_T$($\subset$), $\perp_T>\times_T>0_T>1_T$($\vdash$) and $\times_T>1_T>0_T>\perp_T$($\dashv$).
	
	\begin{minipage}{\textwidth}
		\begin{minipage}[t]{0.45\textwidth}
			\centering
			\makeatletter\def\@captype{table}\makeatother
			\begin{tabular}{c|c c c c}
				$\sqcap$ & $1_T$ & $0_T$ & $\perp_T$ & $\times_T$\\
				\hline
				$1_T$ & $1_T$ & $0_T$ & $\perp_T$ & $\times_T$\\
				$0_T$ & $0_T$ & $0_T$ & $\perp_T$ & $\times_T$\\
				$\perp_T$ & $\perp_T$ & $\perp_T$ & $\perp_T$ & $\times_T$\\
				$\times_T$ & $\times_T$ & $\times_T$ & $\times_T$ & $\times_T$\\
			\end{tabular}
			\caption{Binary Operator $\sqcap$}
		\end{minipage}
		\begin{minipage}[t]{0.45\textwidth}
			\centering
			\makeatletter\def\@captype{table}\makeatother
			\begin{tabular}{c|c c c c}
				$\cup$ & $1_T$ & $0_T$ & $\perp_T$ & $\times_T$\\
				\hline
				$1_T$ & $1_T$ & $0_T$ & $\perp_T$ & $\times_T$\\
				$0_T$ & $0_T$ & $0_T$ & $\perp_T$ & $\times_T$\\
				$\perp_T$ & $\perp_T$ & $\perp_T$ & $\perp_T$ & $\perp_T$\\
				$\times_T$ & $\times_T$ & $\times_T$ & $\perp_T$ & $\times_T$\\
			\end{tabular}
			\caption{Binary Operator $\cup$}
			
		\end{minipage}
	\end{minipage}
	
	\begin{minipage}{\textwidth}
		\begin{minipage}[t]{0.45\textwidth}
			\centering
			\makeatletter\def\@captype{table}\makeatother
			\begin{tabular}{c|c c c c}
				$\cap$ & $1_T$ & $0_T$ & $\perp_T$ & $\times_T$\\
				\hline
				$1_T$ & $1_T$ & $0_T$ & $\perp_T$ & $\times_T$\\
				$0_T$ & $0_T$ & $0_T$ & $0_T$ & $\times_T$\\
				$\perp_T$ & $\perp_T$ & $0_T$ & $\perp_T$ & $\times_T$\\
				$\times_T$ & $\times_T$ & $\times_T$ & $\times_T$ & $\times_T$\\
			\end{tabular}
			\caption{Binary Operator $\cap$}
		\end{minipage}
		\begin{minipage}[t]{0.45\textwidth}
			\centering
			\makeatletter\def\@captype{table}\makeatother
			\begin{tabular}{c|c c c c}
				$\sqsubset$ & $1_T$ & $0_T$ & $\perp_T$ & $\times_T$\\
				\hline
				$1_T$ & $1_T$ & $0_T$ & $\times_T$ & $\perp_T$\\
				$0_T$ & $0_T$ & $0_T$ & $0_T$ & $0_T$\\
				$\perp_T$ & $\perp_T$ & $0_T$ & $\perp_T$ & $\perp_T$\\
				$\times_T$ & $\times_T$ & $0_T$ & $\perp_T$ & $\times_T$\\
			\end{tabular}
			\caption{Binary Operator $\sqsubset$}
			
		\end{minipage}
	\end{minipage}

	\begin{minipage}{\textwidth}
		\begin{minipage}[t]{0.45\textwidth}
			\centering
			\makeatletter\def\@captype{table}\makeatother
			\begin{tabular}{c|c c c c}
				$\wedge$ & $1_T$ & $0_T$ & $\perp_T$ & $\times_T$\\
				\hline
				$1_T$ & $1_T$ & $1_T$ & $\perp_T$ & $\times_T$\\
				$0_T$ & $1_T$ & $0_T$ & $\perp_T$ & $\times_T$\\
				$\perp_T$ & $\perp_T$ & $\perp_T$ & $\perp_T$ & $\times_T$\\
				$\times_T$ & $\times_T$ & $\times_T$ & $\times_T$ & $\times_T$\\
			\end{tabular}
			\caption{Binary Operator $\wedge$}
		\end{minipage}
		\begin{minipage}[t]{0.45\textwidth}
			\centering
			\makeatletter\def\@captype{table}\makeatother
			\begin{tabular}{c|c c c c}
				$\vee$ & $1_T$ & $0_T$ & $\perp_T$ & $\times_T$\\
				\hline
				$1_T$ & $1_T$ & $1_T$ & $1_T$ & $1_T$\\
				$0_T$ & $1_T$ & $0_T$ & $\perp_T$ & $\times_T$\\
				$\perp_T$ & $1_T$ & $\perp_T$ & $\perp_T$ & $\times_T$\\
				$\times_T$ & $1_T$ & $\times_T$ & $\times_T$ & $\times_T$\\
			\end{tabular}
			\caption{Binary Operator $\vee$}
			
		\end{minipage}
	\end{minipage}

	\begin{minipage}{\textwidth}
		\begin{minipage}[t]{0.45\textwidth}
			\centering
			\makeatletter\def\@captype{table}\makeatother
			\begin{tabular}{c|c c c c}
				$\supset$ & $1_T$ & $0_T$ & $\perp_T$ & $\times_T$\\
				\hline
				$1_T$ & $1_T$ & $0_T$ & $\perp_T$ & $1_T$\\
				$0_T$ & $0_T$ & $0_T$ & $\perp_T$ & $0_T$\\
				$\perp_T$ & $\perp_T$ & $\perp_T$ & $\perp_T$ & $\perp_T$\\
				$\times_T$ & $1_T$ & $0_T$ & $\perp_T$ & $\times_T$\\
			\end{tabular}
			\caption{Binary Operator $\supset$}
		\end{minipage}
		\begin{minipage}[t]{0.45\textwidth}
			\centering
			\makeatletter\def\@captype{table}\makeatother
			\begin{tabular}{c|c c c c}
				$\subset$ & $1_T$ & $0_T$ & $\perp_T$ & $\times_T$\\
				\hline
				$1_T$ & $1_T$ & $1_T$ & $1_T$ & $1_T$\\
				$0_T$ & $1_T$ & $0_T$ & $0_T$ & $0_T$\\
				$\perp_T$ & $1_T$ & $0_T$ & $\perp_T$ & $\perp_T$\\
				$\times_T$ & $1_T$ & $0_T$ & $\perp_T$ & $\times_T$\\
			\end{tabular}
			\caption{Binary Operator $\subset$}
			
		\end{minipage}
	\end{minipage}

	\begin{minipage}{\textwidth}
		\begin{minipage}[t]{0.45\textwidth}
			\centering
			\makeatletter\def\@captype{table}\makeatother
			\begin{tabular}{c|c c c c}
				$\vdash$ & $1_T$ & $0_T$ & $\perp_T$ & $\times_T$\\
				\hline
				$1_T$ & $1_T$ & $1_T$ & $1_T$ & $1_T$\\
				$0_T$ & $1_T$ & $0_T$ & $0_T$ & $0_T$\\
				$\perp_T$ & $1_T$ & $0_T$ & $\perp_T$ & $\times_T$\\
				$\times_T$ & $1_T$ & $0_T$ & $\times_T$ & $\times_T$\\
			\end{tabular}
			\caption{Binary Operator $\vdash$}
		\end{minipage}
		\begin{minipage}[t]{0.45\textwidth}
			\centering
			\makeatletter\def\@captype{table}\makeatother
			\begin{tabular}{c|c c c c}
				$\dashv$ & $1_T$ & $0_T$ & $\perp_T$ & $\times_T$\\
				\hline
				$1_T$ & $1_T$ & $0_T$ & $\perp_T$ & $\times_T$\\
				$0_T$ & $0_T$ & $0_T$ & $0_T$ & $\times_T$\\
				$\perp_T$ & $\perp_T$ & $0_T$ & $\perp_T$ & $\times_T$\\
				$\times_T$ & $\times_T$ & $\times_T$ & $\times_T$ & $\times_T$\\
			\end{tabular}
			\caption{Binary Operator $\dashv$}
			
		\end{minipage}
	\end{minipage}

	In addition, the target can be composed of a group of atomic targets          \{\emph{$t_1$}, \emph{$t_2$}...\emph{$t_n$},\}, combined with these operators. For the target expression, the binary operators have the same priority and run the same priority operators from left to right. Binary operators take priority over an unary operator, but operators in brackets should be calculated first. 
	
	\begin{center}
		$\lnot$ ($t_1$ $\wedge$ $t_2$ $\cap$ $t_3$ $\cup$ $t_4$)\\
	\end{center}
	
	\begin{center}
		$\sim$ ($\lnot t_1 \sqcup \sim t_2 \wedge (t_3 \cap t_4)$)\\
	\end{center}
	
	In the first expression above, the calculation should start from the operators in brackets. If there are no brackets, it should start with the unary operator $\lnot$. Since the operators inside of the brackets are binary, they are calculated from left to right, then, calculating $\lnot$ is last. In the second expression, it is obvious that the sub-expression inside of the first bracket should be calculated first. $\lnot t_1$, $\sim t_2$ and $t_3 \cap t_4$ need to be calculated first, as an unary operator and operators in brackets take priority. Then, $\sqcup$ and $\wedge$ combine results of both in brackets, which is taken as $\sim$ in the end.
	
	A solid target in a policy can be represented as constructs of trees with nodes, where leaf nodes are atomic targets and non-leaf nodes are logic gates. We employ a policy tree structure to combine these atomic targets. The lower expression above was illustrated in Figure 4.5. 

	\begin{figure}[thb]
		\begin{minipage}[t]{0.5\linewidth}
			\vspace{-0cm}
			\centering
			\includegraphics[scale=0.35]{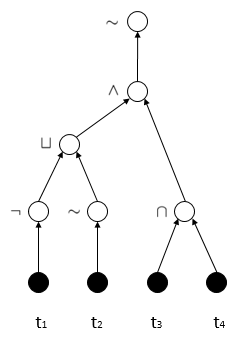}
			\vspace{0.1cm}
			\caption{Target Tree A} \label{fig1}
		\end{minipage}
		\begin{minipage}[t]{0.5\linewidth}
			\vspace{-0cm}
			\centering
			\includegraphics[scale=0.4]{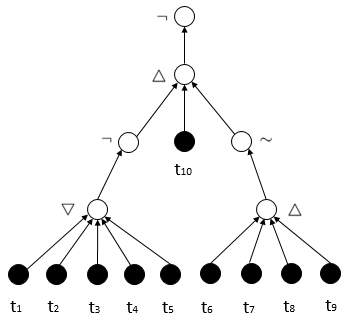}
			\vspace{-0.2cm}
			\caption{Target Tree B} \label{fig1}
		\end{minipage}
	\end{figure}
	
	The result of the target expression indicates whether a policy matches a given query. Only $1_T \in Dec_T$ for target evaluation represents that the policy is applicable. More formally,
	
	-If \emph{t} evaluates to $1_T$, it evaluates \emph{p};\\
	\indent -If \emph{t} evaluates to $0_T$, $\perp_T$ and $\times_T$ it stops further evaluation.
	
	However, to make the expressions more concise, we define u-ary operators, which executes on more than two atomic elements. The $\bf{and}$ and $\bf{or}$ pair for multi-elements are $\bigtriangleup$ and $\bigtriangledown$, corresponding to $\sqcup$ and $\sqcap$ in binary operators. They inherit the same $Dec_T$ priority order as $\sqcup$ and $\sqcap$, which is $1_T$ $>$ $0_T$ $>$ $\perp_T$ $> \times_T$. For $\bigtriangleup$, it selects the highest value among candidates. While $\bigtriangledown$ takes the lowest value among candidates. For instance, $\bigtriangleup$[$1_T$, $0_T$, $\perp_T$, $\times_T$]= $1_T$, $\bigtriangledown$[$1_T$, $0_T$, $\perp_T$, $\times_T$] = $\times_T$. 
	
	Although the u-ary operators can be translated into a series of binary operators, it makes the atomic tree structure briefer and reduces the numbers of operators by combining many binary operators as one. Therefore, to demonstrate the u-ray operators, we provide an example under the tree structure in Figure 4.6. In this example, u-ary operators are used to combine atomic target as: $\bigtriangledown$[$t_1$, $t_2$, $t_3$, $t_4$, $t_5$]; $\bigtriangleup$[$t_6$, $t_7$, $t_8$, $t_9$].

	\subsection{Target Equivalence}
	
	For two targets $t_1$ and $t_2$, if for all requests, the evaluation are always the same, $[\![$ $t_1$ $]\!]$(q)=$[\![$ $t_2$ $]\!]$(q), then $t_1$ and $t_2$ are equivalent targets, $[\![$ $t_1$ $]\!]$ = $[\![$ $t_2$ $]\!]$. We show following priorities of our target operators. 
	
	{\bf Proposition 1} $\forall$ t and t',

	\begin{center}
		$[\![\lnot(\lnot t)]\!]$=$[\![$ $t$ $]\!]$; $[\![\sim(\sim t)]\!]$=$[\![$ $\sim t$ $]\!]$
		
		$[\![\sim(\lnot t)]\!]$ $\neq$ $[\![\lnot(\sim t)]\!]$ 
		
		$[\![\lnot(t\sqcup t')]\!]$ $\neq$ $[\![(\lnot t)\sqcap (\lnot t')]\!]$; $[\![\lnot(t\sqcap t')]\!]$ $\neq$ $[\![(\lnot t)\sqcup (\lnot t')]\!]$
		
		$[\![\sim(t\sqcup t')]\!]$=$[\![(\sim t)\sqcup (\sim t')]\!]$; $[\![\sim(t\sqcap t')]\!]$=$[\![(\sim t)\sqcap (\sim t')]\!]$

	\end{center}

	{\bf Proof} The above equations can be proven by checking the four-valued ``truth" table  defined by operators. 
	
	{\bf Proposition 2} $\forall$ $t_1$, $t_2$ and $t_3$......$t_n$, 
	
	\begin{center}
		$\bigtriangleup$[$t_1$, $t_2$, $t_3$] = $t_1$ $\sqcup$ $t_2$ $\sqcup$ $t_3$;
		
		$\bigtriangledown$[$t_1$, $t_2$, $t_3$] = $t_1$ $\sqcap$ $t_2$ $\sqcap$ $t_3$;
		
		$\bigtriangleup$[$t_1$, $t_2$...... $t_n$] = $t_1$ $\sqcup$ $t_2$ $\sqcup$...... $t_n$;
		
		$\bigtriangledown$[$t_1$, $t_2$...... $t_3$] = $t_1$ $\sqcap$ $t_2$ $\sqcap$...... $t_3$;
		
	\end{center}
	
	{\bf Proof}   The above equations can be proven by checking the four-valued ``truth" table  defined by operators.

	\subsection{On functional Completeness}
	
	We wish the operations we defined to express all functions that users would like to define. Therefore, we prove $\forall \emph{f}$ : $Dec_T^n \rightarrow Dec_T$, n$\in Z^*$, where f can be expressed by at least one combination of $\lnot$, $\sim$, $\star$ and $\cup$. This property is proven by the logic expressions defined by $\lnot$, $\sim$, $\star$ and $\cup$ over $Dec_T$ is functional completeness. We prove this by introducing a theorem from Jobe \cite{Jobe62}.

	{\bf Theorem 1 (Jobe) 1962} The three-valued logic E expressed over the set {1,2,3} and defined by the operators $\bullet$, $E_1$ and $E_2$, given in the figure below, is functionally complete. \\
	
	\begin{minipage}{\textwidth}
		\begin{minipage}[t]{0.45\textwidth}
			\centering
			\makeatletter\def\@captype{table}\makeatother
			\begin{tabular}{c|c c c|c|c}
				$\bullet$ & 3 & 2 & 1 & $E_1$ & $E_2$\\
				\hline
				3 & 3 & 2 & 1 & 3 & 1\\
				2 & 2 & 2 & 1 & 1 & 2\\
				1 & 1 & 1 & 1 & 2 & 3\\
			\end{tabular}
			\caption{Over the Set \{3,2,1\}}
		\end{minipage}
		\begin{minipage}[t]{0.45\textwidth}
			\centering
			\makeatletter\def\@captype{table}\makeatother
			\begin{tabular}{c|c c c c|c|c|c}
				$\bullet$ & $1_T$ & $0_T$ & $\perp_T$ & $\times_T$ & $E_1$ & $E_2$ & $E_3$\\
				\hline
				$1_T$ & $1_T$ & $0_T$ & $\perp_T$ & $\times_T$ & $0_T$ & $1_T$ & $\perp_T$\\
				$0_T$ & $0_T$ & $0_T$ & $\perp_T$ & $\times_T$ & $1_T$ & $\perp_T$ & $\times_T$\\
				$\perp_T$ & $\perp_T$ & $\perp_T$ & $\perp_T$ & $\times_T$ & $\perp_T$ & $0_T$ & $1_T$\\
				$\times_T$ & $\times_T$ & $\times_T$ & $\times_T$ & $1_T$ & $\times_T$ & $\times_T$ & $0_T$\\
			\end{tabular}
			\caption{(Over Set \{$1_T$, $0_T$, $\perp_T$, $\times_T$\}}
			
		\end{minipage}
	\end{minipage}

	{\bf Theorem 2 J-operators and functional completeness.\cite{Jobe62}} In order to present a new and simple proof of functional completeness, we define the operators $J_k(P_1, P_2, ... P_n)^l$, which have the following interpretation. $J_k(P_1, P_2, ... P_n)^l$ represents the truth table of a possible formula of order n which has the truth value k in the $i^{th}$ row and the truth value 1 in all other positions, where (1$\leq$k$\leq$M) and ($1 \leq i \leq M^n$).\\
	
	{\bf Corollary 2} Logic operations $\cup$, $\lnot$, $\sim$, and $\star$ is functionally complete for expressions over four-valued set \{$1_T$, $0_T$, $\perp_T$, $\times_T$\}.\\

	{\bf Proof} Based on the theorem in Jobe's paper\cite{Jobe62}, the operators we tailored for our four-valued set is functionally complete.
	
	\section{Access Control Policy}
	
	When the \emph{target} section is evaluated to be true, this indicates that the request is within the effective range of the policy. In the access control section, we still employ partitions in provenance as \emph{atomic conditions}. The access control policy section determines that the access request can be allowed/denied if the provenance graphs contain certain provenance partitions in the policy. Based on both target section and access control section employing provenance partitions as conditions, the model of atomic condition takes the same form as the atomic target. However, the difference is that each section may take different provenance partitions as conditions. 
	
	Here, we formally define atomic conditions. Let $\{1_P, 0_P, \perp_P, \times_T\}$ $\in Dec_P$ is a decision set;\\
	- p $\in Dec_P$ is an atomic condition, where p is a value in the decision set;\\
	- a provenance partition is an atomic condition;\\
	- (t$\stackrel{\diamond} \rightarrow$p\{tag\}) is a policy; and t is an atomic target, where $\diamond$ is $\prec$ or $\succ$ ($\prec$ and $\succ$ are two approaches to transform atomic \emph{target} as atomic condition);\\ 
	- ($^*t\stackrel{\diamond} \rightarrow$p\{tag\}) is an atomic condition; where $^*$ is a logic unary expression over an atomic target, and $\diamond$ is $\prec$ or $\succ$;\\ 
	- $p_1 \bullet p_2$ which is a conjunction policy of $p_1$ and $p_2$, where $\bullet$ is an operator connection two atomic conditions;\\
	- $pbd_P$ $p$ permit by default policy is a policy p, where $pbd_P$ can be treated as an operator to map a four-valued decision set to a two-valued decision set (Yes or No);\\
	- $dbd_P$ $p$ deny by default policy is a policy p, where $pbd_P$ can be treated as an operator to map a four-valued decision set to a two-valued decision set (Yes or No);\\
	
	Similarly, with the target section, we still utilise a tree structure to organise atomic conditions. The tree structure enables proper visualisation of a policy and supports a flexible and powerful combination of atomic conditions. Even though atomic conditions employ different provenance partitions, they take the same model to extract elements from provenance. Atomic conditions could be provenance partitions already existing in the target section or in newly defined provenance partitions. To be more specific, atomic access control policies could recall conditions in \emph{target} to reduce redundancy. Their conditions could be re-used as a form of transformation which will be introduced in the later part of this section. So it becomes apparent that the atomic access control policies could also define new provenance partitions as conditions. The following two truth tables define how to re-use the transformation of atomic target policies as atomic access control policies. \emph{Atomic targets} are forwardly or adversely referred to a value in the four-valued decision set $\{1_P$, $0_P$, $\perp_P$, $\times_P\}$. Just as the tables shown below, the transformation takes the form as t$\stackrel{\diamond} \rightarrow$p\{tag\}, and the \{tag\}$\in \{1, 0, \times, \perp\}$. When t is referred to p forwardly, the tag $\times$ keeps the results of \emph{t}; \emph{$\perp$} changes $\times_T$ and $\perp_T$ as $\perp_P$; \emph{0} changes $\times_T$ and $\perp_T$ and $0_T$ as $0_P$; \emph{1} increases all the values as $1_P$. On the contrary, \emph{1} for inverse reference keeps values of \emph{t}; \emph{0} decreases $1_T$ as $0_P$; \emph{$\perp$} decreases $1_T$, $0_T$ as $\perp_P$; and $\perp$ downgrades all values of \emph{t} as $\perp_P$.\\
	
	\begin{minipage}{\textwidth}
		\begin{minipage}[t]{0.45\textwidth}
			\centering
			\makeatletter\def\@captype{table}\makeatother
			\begin{tabular}{c|c |c |c| c|}
				$t \prec p$ & $1$ & $0$ & $\perp$ & $\times$\\
				\hline
				$1_T$ & $1_P$ & $1_P$ & $1_P$ & $1_P$\\
				$0_T$ & $1_P$ & $0_P$ & $0_P$ & $0_P$\\
				$\perp_T$ & $1_P$ & $0_P$ & $\perp_P$ & $\perp_P$\\
				$\times_T$ & $1_P$ & $0_P$ & $\perp_P$ & $\times_P$\\
			\end{tabular}
			\caption{$t \prec p$}
		\end{minipage}
		\begin{minipage}[t]{0.45\textwidth}
			\centering
			\makeatletter\def\@captype{table}\makeatother
			\begin{tabular}{c|c |c| c| c|}
				$t \succ p$ & $1$ & $0$ & $\perp$ & $\times$ \\
				\hline
				$1_T$ & $1_P$ & $0_P$ & $\perp_P$ & $\times_P$\\
				$0_T$ & $0_P$ & $0_P$ & $\perp_P$ & $\times_P$\\
				$\perp_T$ & $\perp_P$ & $\perp_P$ & $\perp_P$ & $\times_P$\\
				$\times_P$ & $\times_P$ & $\times_P$ & $\times_P$ & $\times_P$\\
			\end{tabular}
			\caption{$t \succ p$}
		\end{minipage}
	\end{minipage}

	Moreover, to support all the possibility of references for the atomic target, \emph{atomic targets} can be transformed by unary operators before referred to \emph{p}, which takes the form of $^*t\stackrel{\diamond} \rightarrow$p\{tag\}. We show an example of logic expression over \emph{t} in Table 17 as: $\sim(\lnot t)$, where t $\in$ $\{1_T, 0_T, \perp_T, \times_T\}$. 
	
	\begin{minipage}{\textwidth}
		\begin{minipage}[t]{0.45\textwidth}
			\centering
			\makeatletter\def\@captype{table}\makeatother
			\begin{tabular}{c|c}
				t & $\sim(\lnot t)$ \\
				\hline
				$1_T$ & $0_P$ \\
				$0_T$ & $\perp_P$  \\
				$\perp_T$ & $1_P$  \\
				$\times_T$ & $\times_P$ \\
			\end{tabular}
			\caption{t$\rightarrow\sim (\lnot{t})$}
		\end{minipage}
		\begin{minipage}[t]{0.45\textwidth}
			\centering
			\makeatletter\def\@captype{table}\makeatother
			\begin{tabular}{c|c |c |c| c|}
				$\sim (\lnot t) \prec p$ & $1$ & $0$ & $\perp$ & $\times$ \\
				\hline
				$1_T$ & $1_P$ & $0_P$ & $0_P$ & $0_P$ \\
				$0_T$ & $1_P$ & $0_P$ & $\perp_P$ & $\perp_P$ \\
				$\perp_T$ & $1_P$ & $1_P$ & $1_P$ & $1_P$ \\
				$\times_T$ & $1_P$ & $0_P$ & $\perp_P$ & $\times_P$\\
			\end{tabular}
			\caption{$\sim (\lnot t) \stackrel{\prec} \rightarrow$ p}
		\end{minipage}
	\end{minipage}

	\begin{minipage}{\textwidth}
		\begin{minipage}[t]{0.45\textwidth}
			\centering
			\makeatletter\def\@captype{table}\makeatother
			\begin{tabular}{c|c |c| c| c|}
				$\sim (\lnot t) \succ p$ & $1$ & $0$ & $\perp$ & $\times$ \\
				\hline
				$1_T$ & $1_P$ & $0_P$ & $\perp_P$ & $\perp_P$ \\
				$0_T$ & $1_P$ & $1_P$ & $1_P$ & $1_P$ \\
				$\perp_T$ & $1_P$ & $0_P$ & $0_P$ & $0_P$ \\
				$\times_T$ & $1_P$ & $0_P$ & $\perp_P$ & $\times_P$\\
			\end{tabular}
			\caption{$\sim (\lnot t) \stackrel{\succ} \rightarrow$ p}
		\end{minipage}
		\begin{minipage}[t]{0.45\textwidth}
			\centering
			\makeatletter\def\@captype{table}\makeatother
			\begin{tabular}{c|c |c |c}
				t & $\star(\lnot(\star t))$ & $\sim t$ & $\lnot t$  \\
				\hline
				$1_T$ & $1_T$ & $1_T$ & $0_T$ \\
				$0_T$ & $0_T$ & $\perp_T$ & $1_T$ \\
				$\perp_T$ & $\times_T$ & $0_T$ & $\perp_T$\\
				$\times_T$ & $\perp_T$ & $\times_T$ & $\times_T$\\ 
			\end{tabular}
			\caption{All Orders of a ``Four-valued" set by $\lnot$ and $\sim$}
		\end{minipage}
	\end{minipage}

	{\bf Lemma 1} $\forall t \in$ $Dec_T$, $^*t$ $\stackrel{\diamond} \rightarrow$p{tag} supports all the possibilities for (t, p), where $\diamond$ is $\prec$ or $\succ$, $^*$ is unary expressions over $\lnot$ and $\sim$.\\
	
	{\bf Proof} $^*t$ $\stackrel{\diamond} \rightarrow$p is functionally completed, as $^*t$ can express all the order for a four-valued set and $\diamond$ takes both forwardly and reversely. We firstly list all the orders of a four valued set via $\lnot$ and $\sim$.\\

	Therefore, we can conclude the functional completeness of $^*t$ $\stackrel{\diamond} \rightarrow$p.

	A conjunction of policies is a policy. The motivation to define $pbd_P$P and $dbd_P$P is that we need to transfer a four-valued decision set to a two-valued result (permit or deny access). However, it could be processed by combining results of all applicable policies. This transferring can be defined by each policy, or each system can define $pbd_P$ or $dbd_P$ as the default process. A permit by a default policy returns $1_P$ if the output of \emph{P} is 0 or 1; while it returns $\perp_P$ otherwise. On the contrary, a deny by default policy returns $1_P$ if the output of \emph{P} is 1; while it returns $\perp_P$ if the output of \emph{P} is 0 or $\perp_P$. We define the truth table of both as below.\\
	
\begin{table}[ht]
	\makeatletter\def\@captype{table}\makeatother
	\centering
	\begin{tabular}{c|c|c}
		P & $pbd_P$P & $dbd_P$P\\
		\hline
		$1_P$ & $1_P$ & $1_P$\\
		$0_P$ & $1_P$ & $\times_P$ \\
		$\perp_P$ & $1_P$ & $\times_P$ \\
		$\times_P$ & $\times_P$ & $\times_P$ \\
	\end{tabular}
	\caption{Truth Table for $pbd_P$P and $dbd_P$P}
\end{table}
	
	\begin{figure}[ht]
		\centering
		$\begin{array}{cc}
		\includegraphics[width=4cm]{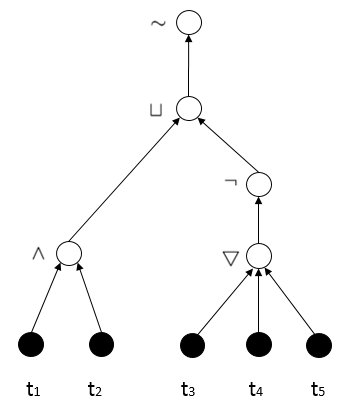} &
		\includegraphics[width=4.5cm]{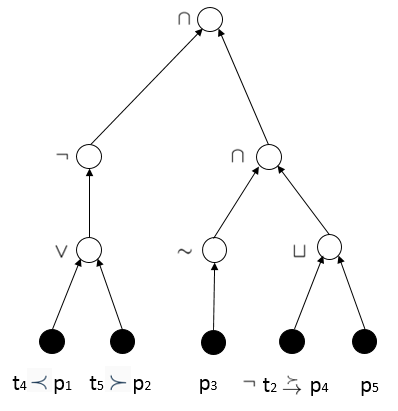} 
		\\
		\end{array}$
		\caption{Example Policy Trees}
		\label{fig:ourlmodels}
	\end{figure}
	\begin{center}
		$\sim$ ($t_1$ $\wedge$ $t_2$) $\cup$ ($\lnot$($\bigtriangleup$[$t_3$, $t_4$, $t_5$]))
	\end{center}
	
	\begin{center}
		$\lnot$ ($t_4\prec p_1$ $\vee$ $t_5\succ p_2$) $\cap$ ($\sim p_3 \sqcup ( (\lnot t_2$  $\stackrel{\succ} \rightarrow$ $p_4) \cap p_5$ )
	\end{center}
	
	In Fig \ref{fig:ourlmodels}, we provide a policy example which includes target policy and access control policy. For this example, atomic target $t_2$ $t_4$, and $t_5$ was assigned as atomic conditions by $t_4 \prec p_1$, $t_5 \succ p_2$, and $\lnot t_2 \stackrel{\succ} \rightarrow p_4$.

	Clearly, the frameworks of the paper titled “PACLP: A Partition-Based Access Control Policy Language for Provenance" and this paper introduce provenance partitions as conditions of policies. However, in provenance access control policies, the decision set is a two-valued. And we defined a four-valued decision set for provenance-based access control policies. Because that provenance-based access control policies mainly depend on attributes of provenance as conditions. The access control policies is to access provenance. Hence, the evaluation for provenance partitions is more fine-grained, where we employ a four-valued decision set in the framework.
	
	\subsection{Policy Operators}
	
	The operators defined in the previous section can be used for access control policy tree structure. In the policy tree structure, logic operations are non-leaf nodes in the policy trees. Here, we introduce two operations inspired by the operators provided by Crampton and Morisset \cite{CramptonM12}. Compared to other operators we proposed, the operators that we define here are aware the orders of inputs. Namely, when the orders of two inputs switch, the output is not always the same. \\

	\begin{minipage}{\textwidth}
		\begin{minipage}[t]{0.45\textwidth}
			\centering
			\makeatletter\def\@captype{table}\makeatother
			\begin{tabular}{c|c |c |c| c}
				$\oplus$ & $1_P$ & $0_P$ & $\perp_P$ & $\times_P$ \\
				\hline
				$1_P$ & $1_P$ & x & $1_P$ & $1_P$\\
				$0_P$ & y & $0_P$ & $0_P$ & $0_P$\\
				$\perp_P$ & $1_P$ & $0_P$ & $\perp_P$ & $\perp_P$\\
				$\times_P$ & $1_P$ & $0_P$ & $\perp_P$ & $\times_P$\\
			\end{tabular}
			\caption{Idemponent}
		\end{minipage}
		\begin{minipage}[t]{0.45\textwidth}
			\centering
			\makeatletter\def\@captype{table}\makeatother
			\begin{tabular}{c|c |c |c| c}
				$\oplus_\cup$ & $1_P$ & $0_P$ & $\perp_P$ & $\times_P$ \\
				\hline
				$1_P$ & $1_P$ & $1_P$ & $1_P$ & $1_P$\\
				$0_P$ & $0_P$ & $0_P$ & $0_P$ & $0_P$\\
				$\perp_P$ & $1_P$ & $0_P$ & $\perp_P$ & $\perp_P$\\
				$\times_P$ & $1_P$ & $0_P$ & $\perp_P$ & $\times_P$\\
			\end{tabular}
			\caption{Idemponent $\oplus_\cup$}
		\end{minipage}
	\end{minipage}

	\begin{minipage}{\textwidth}
		\begin{minipage}[t]{0.45\textwidth}
			\centering
			\makeatletter\def\@captype{table}\makeatother
			\begin{tabular}{c|c |c |c| c}
				$\oplus_\cap$ & $1_P$ & $0_P$ & $\perp_P$ & $\times_P$\\
				\hline
				$1_P$ & $1_P$ & $0_P$ & $1_P$ & $1_P$\\
				$0_P$ & $1_P$ & $0_P$ & $0_P$ & $0_P$\\
				$\perp_P$ & $1_P$ & $0_P$ & $\perp_P$ & $\perp_P$\\
				$\times_P$ & $1_P$ & $0_P$ & $\perp_P$ & $\times_P$\\
			\end{tabular}
			\caption{Idemponent $\oplus_\cap$}
		\end{minipage}
		\begin{minipage}[t]{0.45\textwidth}
			\centering
			\makeatletter\def\@captype{table}\makeatother
			\begin{tabular}{c|c |c| c| c}
				$\triangleright$ & $1_P$ & $0_P$ & $\perp_P$ & $\times_P$\\
				\hline
				$1_P$ & $1_P$ & $1_P$ & $1_P$ & $1_P$\\
				$0_P$ & $0_P$ & $0_P$ & $0_P$ & $0_P$\\
				$\perp_P$ & $1_P$ & $0_P$ & $\perp_P$ & $\perp_P$\\
				$\times_P$ & $1_P$ & $0_P$ & $\perp_P$ & $\times_P$\\
			\end{tabular}
			\caption{First-applicable}
		\end{minipage}
	\end{minipage}

	\begin{table}[ht]
	\makeatletter\def\@captype{table}\makeatother
	
		\centering
		\begin{tabular}{c|c |c| c| c}
			$\triangleleft$ & $1_P$ & $0_P$ & $\perp_P$ & $\times_P$\\
			\hline
			$1_P$ & $1_P$ & $0_P$ & $1_P$ & $1_P$\\
			$0_P$ & $1_P$ & $0_P$ & $0_P$ & $0_P$\\
			$\perp_P$ & $1_P$ & $0_P$ & $\perp_P$ & $\perp_P$\\
			$\times_P$ & $1_P$ & $0_P$ & $\perp_P$ & $\times_P$\\
		\end{tabular}

		\caption{First-applicable}
	\end{table}
	
	We list the truth tables for $\oplus$, $\triangleright$ and $\triangleleft$ as above. In Table 18, the outputs are no longer symmetrical at two sides of the diagonal, where inputs of \{$1_P$, $0_P$\} and \{$0_P$, $1_P$\} differ. For $\oplus$, the priority of four valued sets are $1_P$ = $0_P$ $>$ $\perp_P$ $>$ $\times_P$. Here, we employ $_\cup$ and $_\cap$ to decide x and y when $1_P$ meets $0_P$. In particular, $\oplus_\cup$ takes the former one as the result, while $\oplus_\cap$ takes the latter one as the result. Based on $\oplus$, we can define more operations with the priority order s such as $1_P$ $>$ $0_P$ = $\perp_P$ $>$ $\times_P$, $1_P$ = $\perp_P$ $>$ $0_P$ $>$ $\times_P$ \emph{etc.} 
	
	Similarly, $\triangleright$ is another logic operator that is aware of the order of inputs, and $\triangleright$ emphasises the first input. When the first input is $1_P$ or $0_P$, it covers the second input and was outputted as the result. While, when the first input is $\perp_P$, the second input is the result. $\triangleleft$ effects with the opposite rule by emphasising the second input.
	
	\subsection{A case study}
	
	In order to better deliver the idea, we generate a sample policy and a provenance graph. Thus we output a result of the policy based on the provenance graph.

\noindent \fbox{\parbox{\textwidth}{\ttfamily{
			Atomic Target Policy Section:\\
			Atomic Target (1): Bob uploads a piece of data \\and Alice submits it;\\	
			Atomic Target (2): The data was reviewed on Monday and \\was graded at 30/5/2018;\\
			Access Control Policy Section:\\
			Atomic Condition (1): Bob replaces the data;\\		
			Atomic Condition (2): it was revised after reviewed;\\	
			Atomic Condition (3): it was re-submitted on Wednesday; } }}\\

		\begin{figure*}[ht]
		\centering
		$\begin{array}{cc}
		\includegraphics[width=5cm]{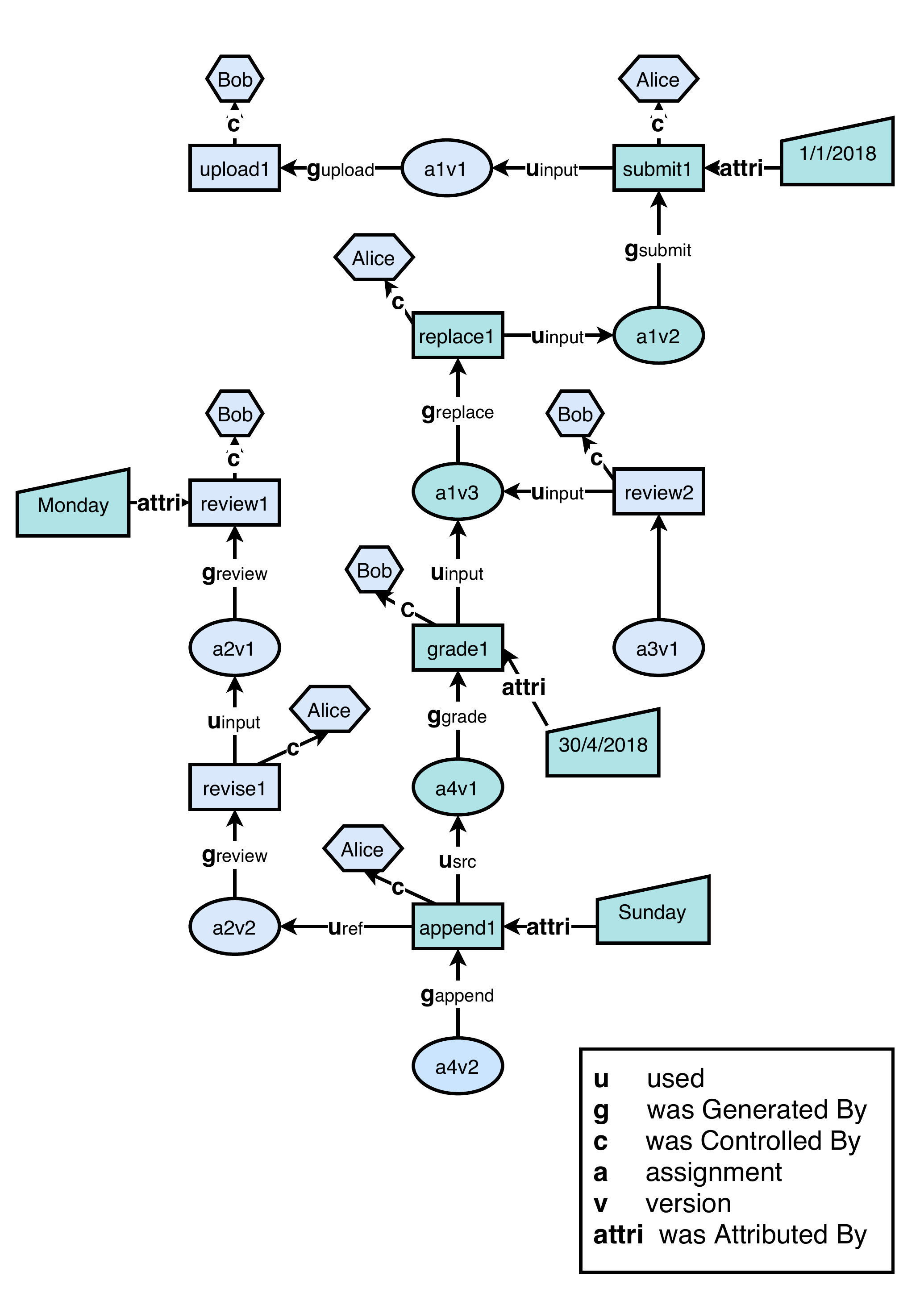} &
		\includegraphics[width=5cm]{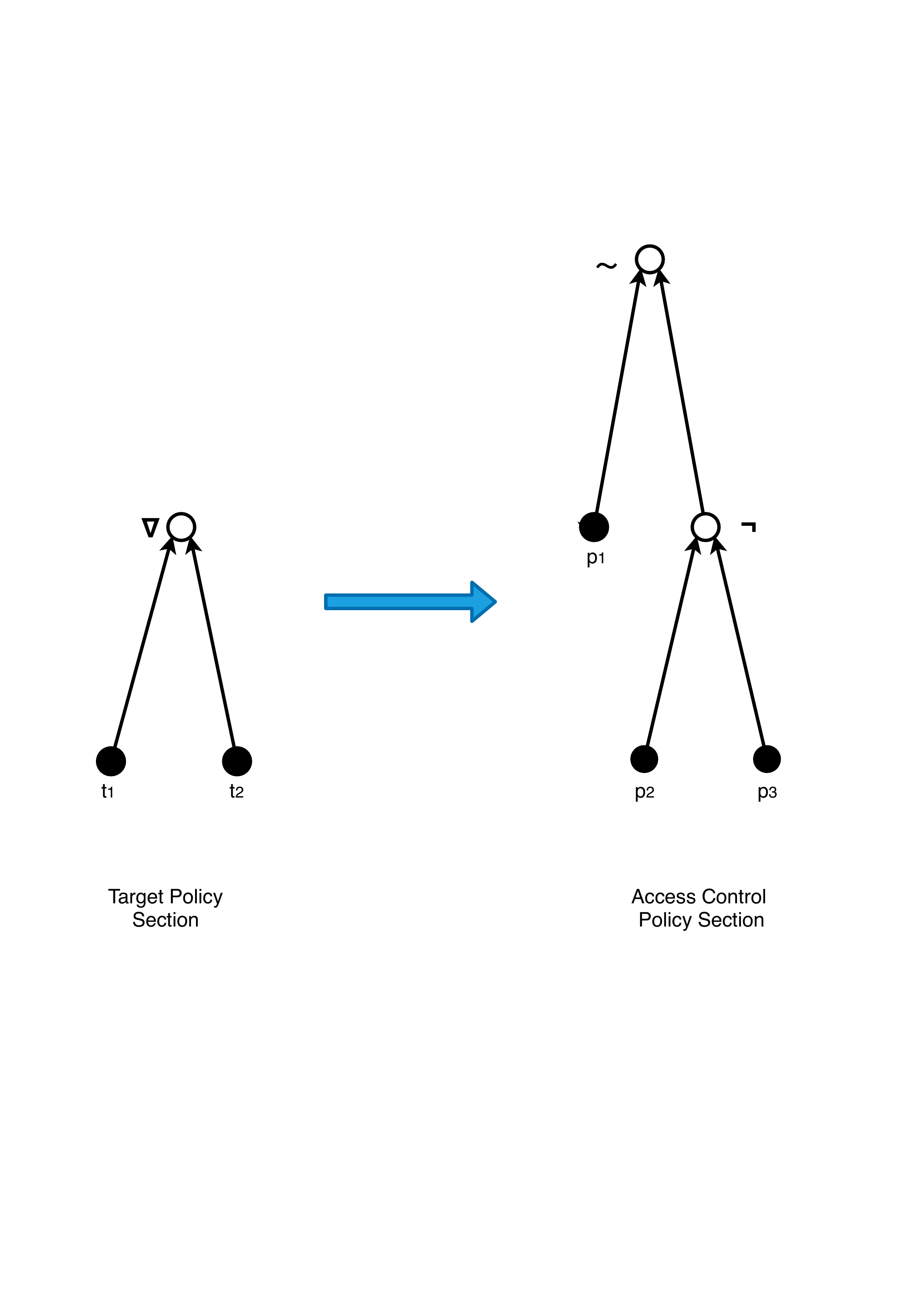} 
		\\
		\mbox{(\textsc{The example provenance graph})} & \mbox{(\textsc{The example policies})}
		\end{array}$
		\caption{A Case Study}
		\label{fig:ourlmodels11}
	\end{figure*}

	A request was sent to access a piece of data in a database, the server retrieves its provenance graph and makes a decision based on the sample policy. Begin with the target section, referring to the provenance graph, we can see Bob uploaded the data ``a1v1" and Alice submitted it. Hence, the results for the atomic target (1) is total match $1_T$. Next, in the provenance graph, the data was reviewed at Monday but graded at 30/5/2018 instead of 30/4/2018, which indicates that the provenance graphs partial match the atomic target (2), where its result is $0_T$. The results of the atomic targets were merged by the operator $\sqcup$. Referring the Table 4.2 of this paper, $1_T \sqcup 0_T$ = $1_T$. It shows that the given provenance graph matches that target section, therefore, we move to the access control policy section. Similarly, as Bob did not replace the data, the result for atomic conditions (1) is $\perp_P$ which merged by the operator $\bigtriangleup$ with outputs of atomic condition (2) $\cap$ atomic condition (3). The final output of this policy is $1_P$. The request to access the data was accepted.

	\section{Conclusion}
	
	In this paper, we propose a fine-grained provenance-based access control policy model by defining atomic condition and policy algebras, which utilise provenance as conditions to determine accessibility of data. Specifically, several types of atomic targets are provided in this paper. Path atomic target takes attributes in provenance as conditions, which implies that if certain operations are executed on the targeted data, these attributes determine that the operation can or cannot access the data. Associated atomic target takes both attributes of provenance and requests as conditions. This indicates that if the requestor has performed certain operations on the targeted data, it can or can not be accessed. Each atomic target takes the form of a string of vertices, where each vertex is defined by a quaternion of attributes.
	
	Moreover, the results for an atomic target of our provenance-based access control policy is one element in a four-valued set. New logic operators to merge the four-valued results are proposed in the paper. The test experiment shows that our model can correctly deal with the request. 

\bibliographystyle{my}
\bibliography{references}

\end{document}